\documentclass[a4paper]{article}

\usepackage{amsmath,amssymb,a4wide}

\newcommand{\pr}[1]{{\mathrm{Pr}{\left({#1}\right)}}}
\newcommand{\until}[1]{\mathrel{U_{#1}}}

\title{Erratum to: Model-checking continuous-time Markov chains by Aziz et al.}

\author{David N. Jansen\thanks{Institute for Computing and Information Sciences, Radboud Universiteit, P.\,O.\@ Box 9010, 6500\,GL Nijmegen, The Netherlands; e-mail: D.Jansen@cs.ru.nl.}}

\begin{document}

\maketitle

\begin{abstract}
        This note corrects a discrepancy
	between the semantics and the algorithm
	of the multiple until operator of CSL,
	like in $\mathit{Pr}_{> 0.0025} (a \until{[1,2]} b \until{[3,4]} c)$,
	of the article:
	Model-checking continuous-time Markov chains
	by Aziz, Sanwal, Singhal and Brayton, TOCL 1(1), July 2000, pp.\@ 162--170.
\end{abstract}





\section{Introduction}

The widely cited article \cite{ASSB00} defines continuous stochastic logic (CSL),
a logic to reason about continuous-time Markov chains,
with a multiple until operator
to write formulas
(with atomic propositions $a$, $b$, and $c$)
like:
$$a \until{[1,2]} b \until{[3,4]} c \quad .$$
The semantics given in the article is:
	\begin{quote}
		A path $\pi$ satisfies $f_1 \until{[a_1,b_1]} f_2 \until{[a_2,b_2]} \cdots \until{[a_{n-1}, b_{n-1}]} f_n$
		``if and only if there exist real numbers $t_1, \ldots, t_{n-1}$
		such that for each integer in $[1, n]$ we have
		$(a_i \leq t_i \leq b_i) \wedge \linebreak (\forall t' \in [t_{i-1}, t_i]%
		) (\pi(t) \text{'' satisfies $f_i$``})$,
		where $t_{-1}$ is defined to be $0$ for notational convenience.''
	\end{quote}
This definition uses the undefined variables $t$, $t_0$, $i$, $a_n$, $b_n$, and $t_n$
(while it defines the unused variable~$t_{-1}$),
and it seems to require
that $\pi(t_i)$ satisfy $f_i \wedge f_{i+1}$.
Obviously, the authors meant something like:
\begin{quote}
	A path $\pi$ satisfies $f_1 \until{[a_1,b_1]} f_2 \until{[a_2,b_2]} \cdots \until{[a_{n-1}, b_{n-1}]} f_n$
	if and only if
	there exist real numbers $0 < t_1 < t_2 < \cdots < t_{n-1}$
	such that for each integer $i$ in $[1, n-1]$
	we have $(a_i \leq t_i \leq b_i)
	\wedge (\forall t' \in [t_{i-1}, t_i)) (\pi(t') \text{ satisfies } f_i)$,
	where $t_{0}$ is defined to be $0$ for notational convenience,
	and additionally $\pi(t_{n-1})$ satisfies $f_n$.
\end{quote}
However, the implementation,
i.\,e.\@ the algorithm that estimates the probability of this until operator,
uses another semantics implicitly, namely the following:
\begin{quote}
	A path $\pi$ satisfies $f_1 \until{[a_1,b_1]} f_2 \until{[a_2,b_2]} \cdots \until{[a_{n-1}, b_{n-1}]} f_n$
	if and only if
	for each integer $i$ in $[1, n-1]$
	we have
	$(\forall t' \in [b_{i-1}, a_i]\,) (\pi(t') \text{ satisfies } f_i)
	\wedge (\forall t' \in (a_i, b_i) \, ) (\pi(t') \text{ satisfies } \linebreak f_i \vee f_{i+1})$,
	where $b_0$ is defined to be $0$ for notational convenience,
	and additionally $\pi(b_{n-1})$ satisfies $f_n$.
\end{quote}
The implementation allows to switch back and forth
between states satisfying $f_i \wedge \neg f_{i+1}$
and states satisfying $\neg f_i \wedge f_{i+1}$,
and it requires to stay in a $f_n$-state longer than the semantics.

The present article exhibits the error and shows how it can be corrected.
In the remainder of the article, we will assume
that the intervals do not overlap,
i.\,e.\@, that $b_i < a_{i+1}$ for all $i = 1, 2, \ldots, n-2$.

\section{Example}

For example, consider the Markov chain drawn in Figure~\ref{fig.error.example}.
The probability that a path satisfies the formula $g = a \until{[1,2]} b \until{[3,4]} c$
can be calculated as the product of a few Poisson probabilities:
\begin{multline*}
\pr{0 \text{ transitions during time } [0,1)} \cdot
\pr{1 \text{ transition during time } [1,2]} \cdot
\tfrac{1}{2} \cdot \\
\cdot \pr{0 \text{ transitions during time } (2,3)} \cdot
\pr{> 0 \text{ transition(s) during time } [3,4]} = \\
= \frac{2^0 e^{-2}}{0!} \cdot
\frac{2^1 e^{-2}}{1!} \cdot
\tfrac{1}{2} \cdot
\frac{2^0 e^{-2}}{0!} \cdot
\left(1 - \frac{2^0 e^{-2}}{0!}\right) =
\tfrac{2}{2} e^{-6}(1-e^{-2}) \approx 0.00214
\end{multline*}
However, \cite{ASSB00}'s algorithm does the following calculations:
The probability of the formula $g$ is \\
\begin{equation}
\mu^1(g) =
\left( 1, 0, 0, 0 \right)
P_{a}(1) I_{a}
P_{a \vee b}(2-1) I_{b}
P_{b}(3-2) I_{b}
P_{b \vee c}(4-3) I_{c}
\begin{pmatrix} 1 \\ 1 \\ 1 \\ 1 \end{pmatrix}
\label{formula.example}
\end{equation}
where $P_{f}(t)$ is the transition probability matrix of the Markov chain with alls $\neg f$-states changed to absorbing states, for time interval $t$; and $I_{f}$ is the diagonal matrix with entries $1$ for $f$-states and $0$ for $\neg f$-states.
For example,
\begin{figure}[tbp]
	\setlength{\unitlength}{0.3cm}
	\center{
	\begin{picture}(30,5.2)(-3,-2.6)
		\put( 0,0){\raisebox{-0.5\height}{\makebox[0pt]{1}}\circle{3}}
		\put( 8,0){\raisebox{-0.5\height}{\makebox[0pt]{2}}\circle{3}}
		\put(16,0){\raisebox{-0.5\height}{\makebox[0pt]{3}}\circle{3}}
		\put(24,0){\raisebox{-0.5\height}{\makebox[0pt]{4}}\circle{3}}
		\put( 0,-2.6){\makebox[0pt]{$a$}}
		\put( 8,-2.6){\makebox[0pt]{$b$}}
		\put(16,-2.6){\makebox[0pt]{$c$}}
		\put(-2.5,0){\line(1,0){0.8}}	\put(-1.5,0){\vector(1,0){0}}
		\qbezier(1.3,0.75)(4,2.1)(6.5,0.85)	\put(6.7,0.75){\vector(2,-1){0}}
		\put(4,2){\makebox[0pt]{2}}
		\qbezier(6.7,-0.75)(4,-2.1)(1.5,-0.85)	\put(1.3,-0.75){\vector(-2,1){0}}
		\put(4,-2){\raisebox{-\height}{\makebox[0pt]{1}}}
		\put(9.5,0){\line(1,0){4.8}}	\put(14.5,0){\vector(1,0){0}}
		\put(12,0.3){\makebox[0pt]{1}}
		\put(17.5,0){\line(1,0){4.8}}	\put(22.5,0){\vector(1,0){0}}
		\put(20,0.3){\makebox[0pt]{2}}
	\end{picture}
	}
	\caption{Example Markov chain\label{fig.error.example}}
\end{figure}
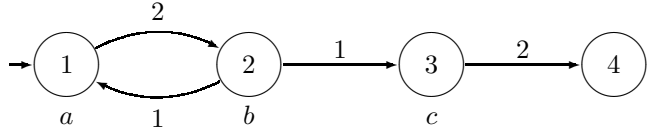
\begin{align*}
P_{a}(t) & =
	\exp \begin{pmatrix}
		-2t & 2t & 0 & 0 \\
		 0  & 0  & 0 & 0 \\
		 0  & 0  & 0 & 0 \\
		 0  & 0  & 0 & 0 \end{pmatrix}
  &
I_{b} & = \begin{pmatrix}
		0 & 0 & 0 & 0 \\
		0 & 1 & 0 & 0 \\
		0 & 0 & 0 & 0 \\
		0 & 0 & 0 & 0 \end{pmatrix}
\end{align*}
Multiplying all these matrices as indicated in Formula~(\ref{formula.example}) produces the outcome:
$$\mu^1(g) = 
\tfrac{1}{2}e^{-8}(e^{\sqrt{2}} - e^{-\sqrt{2}})\sqrt{2}
\approx 0.000918$$
which is less than half the actual value.

\section{First problem: final transition}

A problem arises with paths
that enter state 4 during the time interval $(3,4]$.
These paths are counted as non-satisfying by the algorithm
(Formula~(\ref{formula.example}) only counts the paths
that are in a $c$-state at time 4.),
while they have passed through a $c$-state timely
and actually may satisfy $g$.

To solve this problem,
$P_{b \vee c}$ should be replaced by a matrix
based on a Markov chain
where \emph{additionally} all $c$-states have been made absorbing,
so that a path entering an $c$-state stays there until time $4$.
This is basically the same transformation as described by \cite{BHHK03}
for simple until formulas $f_1 \until{[a_1, b_1]} f_2$:
Make all states except the $f_1 \wedge \neg f_2$-states absorbing.
In our example, we have to replace the factor $P_{b \vee c}(t)$ of Formula~(\ref{formula.example}) by:
$$P_{b \wedge \neg c}(t) =
	\exp \begin{pmatrix}
		0 &  0  & 0 & 0 \\
		t & -2t & t & 0 \\
		0 &  0  & 0 & 0 \\
		0 &  0  & 0 & 0 \end{pmatrix}$$
and so, the calculated probability becomes
\begin{multline}
\mu^1(g) =
\left( 1, 0, 0, 0 \right)
P_{a}(1) I_{a}
P_{a \vee b}(2-1) I_{b}
P_{b}(3-2) I_{b}
P_{b \wedge \neg c}(4-3) I_{c}
\begin{pmatrix} 1 \\ 1 \\ 1 \\ 1 \end{pmatrix} = \\
= \tfrac{1}{4} e^{-6}(1-e^{-2})(e^{\sqrt{2}} - e^{-\sqrt{2}}) \sqrt{2}
\approx 0.00293
\label{formula.example.partly.corrected}
\end{multline}
which unfortunately is still wrong.

\section{Second problem: intermediary transitions}

The remaining discrepancy after the first correction reveals another problem.
In the example, there is a transition from $b$-state 2 back to $a$-state 1.
According to Formula~(\ref{formula.example}), the path $1 \xrightarrow{t=1} 2 \xrightarrow{t=1.5} 1 \xrightarrow{t=1.8} 2 \xrightarrow{t=3.2} 3$
is counted as a path satisfying $g$,
as it continuously satisfies $a \vee b$ during the interval $(1,2)$.
However, the semantics requires
that one choose when $t_1 \in [1,2]$ has come.
This must happen upon entering state 2 (a $b \wedge \neg a$-state) at the latest,
so $t_1 = 1$.
After $t_1$, one is no longer allowed to enter $a \wedge \neg b$-states,
so the transition $2 \xrightarrow{t=1.5} 1$ is forbidden.

\subsection{Wrong correction}

We could try to correct this problem by deleting transitions from $b$-states to $a$-states in the example.
This, however, gives rise to two new problems:
\begin{enumerate}
\item	The exit rate of state 2 (in the Markov chain of Figure~\ref{fig.error.example}) would change.
\item	What about $a \wedge b$-states?
	These states are allowed both before and after $t_1$.
	If $t_1$ has passed,
	switching back and forth between $a \wedge b$-states
	and $b \wedge \neg a$-states should be allowed,
	while it should be counted as an error to enter an $a \wedge \neg b$-state.
	Before $t_1$, the opposite condition holds.
	It is impossible to make a subset of the states absorbing
	in a way that satisfies both conditions.

\end{enumerate}

\subsection{Better correction}

To solve the problem mentioned above,
I propose to add extra states to the Markov chain for $P_{a \vee b}$:
Introduce a second copy $s'$ of each $a$-state $s$
that has a $b$-predecessor state.
One copy ($s$) stands for ``$t_1$ (possibly) has not yet passed'' and the second ($s'$) for ``$t_1$ has passed definitely''.
So, transitions from $b \wedge \neg a$-states to $s$ are deflected to $s'$,
and if $s$ satisfies $a \wedge \neg b$,
then $s'$ is an error state
and is rendered absorbing.

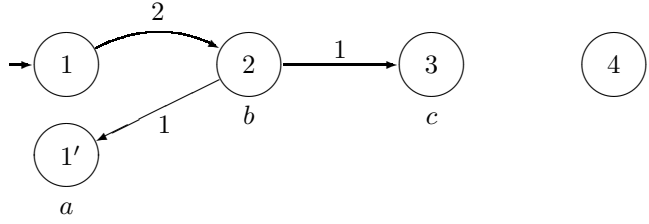
\begin{figure}[tbp]
	\setlength{\unitlength}{0.3cm}
	\center{
	\begin{picture}(30,9.2)(-3,-6.6)
		\put( 0,0){\raisebox{-0.5\height}{\makebox[0pt]{1}}\circle{3}}
		\put( 0,-4){\raisebox{-0.5\height}{\makebox[0pt]{1\makebox[0pt][l]{$'$}}}\circle{3}}
		\put( 8,0){\raisebox{-0.5\height}{\makebox[0pt]{2}}\circle{3}}
		\put(16,0){\raisebox{-0.5\height}{\makebox[0pt]{3}}\circle{3}}
		\put(24,0){\raisebox{-0.5\height}{\makebox[0pt]{4}}\circle{3}}
		\put( 0,-6.6){\makebox[0pt]{$a$}}
		\put( 8,-2.6){\makebox[0pt]{$b$}}
		\put(16,-2.6){\makebox[0pt]{$c$}}
		\put(-2.5,0){\line(1,0){0.8}}	\put(-1.5,0){\vector(1,0){0}}
		\qbezier(1.3,0.75)(4,2.1)(6.5,0.85)	\put(6.7,0.75){\vector(2,-1){0}}
		\put(4,2){\makebox[0pt]{2}}
		\put(6.7,-0.65){\line(-2,-1){5.2}}	\put(1.3,-3.35){\vector(-2,-1){0}}
		\put(4,-3){1}
		\put(9.5,0){\line(1,0){4.8}}	\put(14.5,0){\vector(1,0){0}}
		\put(12,0.3){\makebox[0pt]{1}}
	\end{picture}
	}
	\caption{Modified example Markov chain for $P_{a \vee b}$ from Figure~\ref{fig.error.example}\label{fig.correct.example}}
\end{figure}
In our example,
we have to replace the factor $P_{a \vee b}(t)$ of Formula~(\ref{formula.example.partly.corrected})
by the transition probability matrix of the Markov chain shown in Figure~\ref{fig.correct.example},
denoted by $P'_{a \vee b}(1)$
(I treat state $1'$ as fifth state):
$$P'_{a \vee b}(t) =
	\exp \begin{pmatrix}
		-2t &  2t & 0 & 0 & 0 \\
		 0  & -2t & t & 0 & t \\
		 0  &  0  & 0 & 0 & 0 \\
		 0  &  0  & 0 & 0 & 0 \\
		 0  &  0  & 0 & 0 & 0 \end{pmatrix}
$$
and so, the calculated probability becomes
\pagebreak[0]
\begin{multline*}
\mu^1(g) =
\left( 1, 0, 0, 0 \right)
P_{a}(1) I'_{a}
P'_{a \vee b}(2-1) I''_{b}
P_{b}(3-2) I_{b}
P_{b \wedge \neg c}(4-3) I_{c}
\begin{pmatrix} 1 \\ 1 \\ 1 \\ 1 \end{pmatrix} = \\
= e^{-6}(1-e^{-2}) \approx 0.00214
\end{multline*}
which is the correct answer.

In this formula, I also modified $I'_{a}$ and $I''_{b}$ to include the transformation between the four- and five-state-Markov chains:
\pagebreak[0]
$$I'_{a} = \begin{pmatrix}
		1 & 0 & 0 & 0 & 0 \\
		0 & 0 & 0 & 0 & 0 \\
		0 & 0 & 0 & 0 & 0 \\
		0 & 0 & 0 & 0 & 0 \end{pmatrix}
	\qquad \qquad
I''_{b} = \begin{pmatrix}
		0 & 0 & 0 & 0 \\
		0 & 1 & 0 & 0 \\
		0 & 0 & 0 & 0 \\
		0 & 0 & 0 & 0 \\
		0 & 0 & 0 & 0 \end{pmatrix}
$$

\section{General formulation of the corrected semantics}

When we extend the above corrections to general until formulas,
we get the following basis for an algorithm
to compute the probability of until formulas:
\begin{quote}
	The probability of a formula of the form
	$$g := f_1 \until{[a_1,b_1]} f_2 \until{[a_2,b_2]} f_3 \ldots f_{n-1} \until{[a_{n-1},b_{n-1}]} f_n \quad ,$$
	where the intervals do not overlap,
	is given by
	\begin{multline}
	\mu^s(g) = \pi_s \cdot
		P_{f_1}(a_1) \cdot
		I'_{f_1} \cdot
		P'_{f_1 \vee f_2}(b_1 - a_1) \cdot
		I''_{f_2} \cdot
		P_{f_2}(a_2 - b_1) \cdot \\
		I'_{f_2} \cdot
		P'_{f_2 \vee f_3}(b_2 - a_2) \cdot
		I''_{f_3} \cdot
		P_{f_3}(a_3-b_2) \cdots \\
		I'_{f_{n-2}} \cdot
		P'_{f_{n-2} \vee f_{n-1}}(b_{n-2} - a_{n-2}) \cdot
		I''_{f_{n-1}} \cdot
		P_{f_{n-1}}(a_{n-1} - b_{n-2}) \cdot \\
		I_{f_{n-1}} \cdot
		P_{f_{n-1} \wedge \neg f_n}(b_{n-1} - a_{n-1}) \cdot
		I_{f_n} \cdot
		\begin{pmatrix} 1 \\ \vdots \\ 1 \end{pmatrix}
		\label{formula.correct.product}
	\end{multline}
	where $P_{f}(t)$ is the transition probability matrix for time $t$
	corresponding to the Markov chain where all $\neg f$-states are made absorbing;
	$\pi_s$ is the starting probability distribution
	(which in our case has unity for state $s$ and zeroes otherwise);
	$P'_{f_i \vee f_{i+1}}(t)$ is the transition probability matrix for time $t$
	based on the extended Markov chain (details follow);
	and $I'_{f_i}$ and $I''_{f_{i+1}}$ are the matrices
	that map $f_i$-states and $f_{i+1}$-states, respectively, to and from the extended Markov chain (details follow).
\end{quote}
The following table shows which transitions the extended Markov chain contains,
i.\,e.\@ the Markov chain to base $P'_{f_i \vee f_{i+1}}(t)$ upon.
As mentioned above,
the states are the same as the original Markov chain
with an additional copy $s'$ of each $f_{i}$-state $s$
that has a $f_{i+1}$-predecessor.

\pagebreak

If the original Markov chain contains a transition $s \xrightarrow{\lambda} t$, then the extended Markov chain contains the following transition(s):
\begin{center}
\begin{tabular}{|l|c|c|c|c|}
\hline
				 & \multicolumn{2}{c|}{$s'$ is added} & \multicolumn{2}{c|}{$s'$ is not added} \\
				 & \raisebox{-6pt}[0pt][0pt]{$t'$ is added} & \raisebox{-6pt}[0pt][0pt]{$t'$ is not added} & $t'$ is & $t'$ is not \\
				 & & & added & added \\
\hline
$s \models f_i \wedge \neg f_{i+1}$	 & \multicolumn{2}{c|}{$s \xrightarrow{\lambda} t$; $s'$ is absorbing}	 & \multicolumn{2}{c|}{$s \xrightarrow{\lambda} t$} \\
\hline
$s \models f_i \wedge f_{i+1}$		 & $s \xrightarrow{\lambda} t$ and $s' \xrightarrow{\lambda} t'$ & $s \xrightarrow{\lambda} t$ and $s' \xrightarrow{\lambda} t$ & \multicolumn{2}{c|}{$s \xrightarrow{\lambda} t$} \\
\hline
$s \models \neg f_i \wedge f_{i+1}$	 & \multicolumn{2}{c|}{\raisebox{-6pt}[0pt][0pt]{impossible}} & $s \xrightarrow{\lambda} t'$ & $s \xrightarrow{\lambda} t$ \\
\cline{1-1}\cline{4-5}
$s \models \neg f_i \wedge \neg f_{i+1}$ & \multicolumn{2}{c|}{} & \multicolumn{2}{c|}{$s$ is absorbing} \\
\hline
\end{tabular}
\end{center}

\noindent
$I'_{f_i}$ and $I''_{f_{i+1}}$ can be used to convert the probability vectors to and from the extended Markov chain.
At time $a_i$, all probability mass should go into the first copy $s$ of a state
which has a second copy $s'$, so
$$(I'_{f_i})_{st} = \begin{cases} 1 & \text{if $s=t$ and $s$ satisfies $f_i$} \\
				0 & \text{otherwise} \end{cases}$$
Both copies of $f_i \wedge f_{i+1}$-states are allowed at time $b_i$
(the latest possible $t_i$),
so their probabilities should be added in the modified $I''_{f_{i+1}}$:
$$(I''_{f_{i+1}})_{st} = \begin{cases} 1 & \text{if $s=t$ or $s=t'$ and $t$ satisfies $f_{i+1}$} \\
					0 & \text{otherwise} \end{cases}$$

\subsection{Correctness}

To convince ourselves
that the product in Formula~(\ref{formula.correct.product}) corresponds closely to the semantics
given at the beginning of the article,
let us look at some of its factors.
\begin{itemize}
\item	$P_{a_{n-1}} := I_{f_{n-1}} \cdot P_{f_{n-1} \wedge \neg f_n}(b_{n-1} - a_{n-1}) \cdot I_{f_n} \cdot \left. \mathbf{1} \right\rvert$
	can be seen as a vector of probabilities:
	$P_{a_{n-1}}(s)$ is the probability
	that one gets a path
	for which there exists a $t_{n-1} \in (a_{n-1}, b_{n-1}]$
	such that the path is in $f_{n-1}$-states during the time interval $[a_{n-1}, t_{n-1})$
	and it is in a $f_n$-state at time $t_n$,
	under the condition that it is in state $s$ at time $a_{n-1}$,
	as shown by \cite{BHHK03}.
	(Note that $t_{n-1} \gneqq a_{n-1}$
	as the path has to be in a $f_{n-1}$-state at time $a_{n-1}$.
	This is not a relevant difference
	as the paths with $t_{n-1} = a_{n-1}$ form a set that has probability $0$.)

\item	Let $M_{b_{n-2}} := I_{f_{n-1}} \cdot P_{f_{n-1}}(a_{n-1} - b_{n-2}) \cdot I_{f_{n-1}}$.
	Then, $M_{b_{n-2}}(s,t)$ is the probability
	that a path is in $f_{n-1}$-states during the time interval $[b_{n-2}, a_{n-1}]$
	and ends in state $t$ at time $a_{n-1}$,
	under the condition that it is in state $s$ at time $b_{n-2}$.

	Let $P_{b_{n-2}} := M_{b_{n-2}} \cdot P_{a_{n-1}}$.
	So, $P_{b_{n-2}}(s)$ is the probability
	that one gets a path
	for which there exists a $t_{n-1} \in (a_{n-1}, b_{n-1}]$
	such that the path is in $f_{n-1}$-states during the time interval $[b_{n-2}, t_{n-1})$
	and it is in a $f_n$-state at time $t_n$,
	under the condition that it is in state $s$ at time $b_{n-2}$.

\item	Let $M_{a_{n-2}} := I'_{f_{n-2}} \cdot P'_{f_{n-2} \vee f_{n-1}}(b_{n-2} - a_{n-2}) \cdot I''_{f_{n-1}}$.
	Then, $M_{a_{n-2}}(s,t)$ is the probability
	that one gets a path
	for which there exists a $t_{n-2} \in (a_{n-2}, b_{n-2}]$
	such that the path is in $f_{n-2}$-states during the time interval $[a_{n-2}, t_{n-2})$,
	it is in $f_{n-1}$-states during the time interval $[t_{n-2}, b_{n-2}]$
	and is in state $t$ at time $b_{n-2}$,
	under the condition that the path is in state $s$ at time $a_{n-2}$.

	Let $P_{a_{n-2}} := M_{a_{n-2}} \cdot P_{b_{n-2}}$.
	(Note that $I''_{f_{n-1}} \cdot I_{f_{n-1}} = I''_{f_{n-1}}$.)
	So, $P_{a_{n-2}}(s)$ is the probability
	that one gets a path
	for which there exist $t_{n-2} \in (a_{n-2}, b_{n-2}]$
	and $t_{n-1} \in (a_{n-1}, b_{n-1}]$
	such that the path is in $f_{n-2}$-states during the time interval $[a_{n-2}, t_{n-2})$,
	it is in $f_{n-1}$-states during the time interval $[t_{n-2}, t_{n-1})$
	and it is in a $f_n$-state at time $t_n$,
	under the condition that it is in state $s$ at time $a_{n-2}$.

\item	etc.

\item	Finally, let $P_0 := M_0 \cdot P_{a_1}$,
	and $P_0(s)$ is the probability
	that one gets a path
	for which there exist $t_1 \in (a_1, b_1], \ldots, t_{n-1} \in (a_{n-1}, b_{n-1}]$
	such that the path is in $f_1$-states during the time interval $[0, t_1)$,
	it is in $f_2$-states during the time interval $[t_1, t_2)$,
	\ldots,
	it is in $f_{n-1}$-states during the time interval $[t_{n-2}, t_{n-1})$
	and it is in a $f_n$-state at time $t_n$,
	under the condition that it is in state $s$ at time $0$.
\end{itemize}
The first factor in the product, $\pi_s$, serves to uncondition on the initial state.
So, overall, the product $\pi_s \cdot P_0$ calculates the probability that a path satisfies the semantics given.

\section{Concluding remarks}

The correction presented above provides a calculation principle
to find the probability that a path satisfies an until formula,
corresponding closely to the intended semantics as given in the introduction.
It does no longer require to stay in $f_n$-states overly long;
it does no longer allow to switch back and forth between $f_{i-1}$- and $f_i$-states too often.

The present note passes over a choice that one has if intervals overlap:
Would $a \until{[1,3]} b \until{[2,4]} c$ be satisfied by a path
that jumped from an $a \wedge \neg b$-state to a $c \wedge \neg b$-state
in the interval $[2,3]$,
i.\,e.\@ a path with $t_1 = t_2$?
\cite{ASSB00}'s remarks about overlapping intervals suggest
they choose to forbid such paths,
and my formulation of the semantics is aligned thereto;
however, in some cases it may be more intuitive
to allow them.
It is possible to solve this discrepancy
by adding the probability of $a \until{[2,3]} c$
if desired.
\cite{BHHK03} choose the other way for simple until formulas: they consider $f_2$-states as satisfying $f_1 \until{[0,b_1]} f_2$.

The main goal of \cite{ASSB00} was to prove decidability of CSL model checking.
This erratum does not invalidate their proof idea;
it only requires to fill in slightly different matrices in some proof parts,
but the main argument -- namely, that basic operations on (matrices containing) algebraic numbers produce (matrices containing) algebraic numbers again -- remains valid for the modified matrices.
So, it still holds up that CSL model checking is decidable.

\paragraph*{Acknowledgement.}
Most of the above matrix calculations have been performed using Maple.

\bibliographystyle{plain}
\bibliography{Aziz-erratum}

\end{document}